\documentstyle[twocolumn,prl,aps,epsf]{revtex} 
\begin{document} 
\title{Temperature dependent scattering rates at the Fermi surface of Optimally 
Doped $Bi_2Sr_2CaCu_2O_{8+\delta}$} 
\author{T. Valla$^1$, A. V. Fedorov$^1$, P. D. Johnson$^1$, Q. Li$^2$, G. D. 
Gu$^3$ and N. Koshizuka$^4$}
\address{$^1$ Department of Physics, Brookhaven National Laboratory, Upton, NY, 
11973-5000}
\address{$^2$ Division of Materials Sciences, Brookhaven National Laboratory, 
Upton, NY, 11973-5000}
\address{$^3$ School of Physics, The University of New South Wales, P.O. Box 1, 
Kensington, NSW, Australia   2033}
\address{$^4$ Superconductivity Research Laboratory, ISTEC, 10-13, Shinonome I-
chrome, Koto-ku, Tokyo 135, Japan}

\address{ {\em \bigskip \begin{quote}
For optimally doped $Bi_2Sr_2CaCu_2O_{8+\delta}$, scattering rates in the normal 
state are found to have a linear temperature dependence over most of the Fermi 
surface. In the immediate vicinity of the $(\pi,0)$ point, the scattering rates 
are nearly constant in the normal state, consistent with models in which 
scattering at this point determines the $c$-axis transport. In the 
superconducting state, the scattering rates away from the nodal direction appear 
to level off and become temperature-independent.
\end{quote}}}
\maketitle
Obtaining a detailed understanding of the microscopic mechanism that results in 
high temperature superconductivity remains one of the fundamental challenges of 
condensed matter physics. Angle-resolved photoemission 
(ARPES) has provided a number of important insights into these materials; the 
mapping of the Fermi surface \cite{1}, the identification of $d$-wave symmetry 
of the order parameter in the superconducting state \cite{2}, and the detection 
of a pseudogap for underdoped compounds above the transition temperature $T_C$ 
\cite{3,4}. Recent advances in instrumentation have allowed ARPES to be 
used for studies of the single particle self-energy, $\Sigma$, a 
quantity that reflects fundamental interactions in a many-body system. The real 
part of the self-energy corresponds to the shift in energy, and the imaginary 
part represents the scattering rate (inverse lifetime) due to the interaction. 
Recently, an ARPES study has provided clear evidence that in optimally doped Bi 
2212, in the direction corresponding to the superconducting gap node, the 
imaginary part of the self-energy has a linear temperature dependence, 
independent of binding energy, for small energies, and a linear energy 
dependence, independent of temperature, for large binding energies \cite{5}. 
This behavior is embodied in the marginal Fermi liquid model, which is 
characterized by a lack of quasiparticles at the Fermi energy \cite{6}. Such 
behavior, found at temperatures above and below the superconducting transition 
temperature, is quantum critical in character \cite{5}.  The measured 
temperature dependence of the self-energy and related momentum spread in 
photoemission is reminiscent of the temperature dependence of the resistivity; 
linear in $T$ with negligible zero-temperature intersect \cite{7}. As such, it 
is clearly of interest to determine over what fraction of the Fermi surface the 
linearity of the momentum spread holds.  In the present paper find that it exists 
over most of the Fermi surface ($\approx 70$\%) in the normal state, with a 
slope that is approximately constant for that portion of the Fermi surface. 
However, as one moves away from the nodal point, there is an increase in the 
temperature-independent offset. Only in the immediate vicinity of the $(\pi,0)$ 
point, the momentum widths appear to level off, with no temperature dependence. 
These observations suggest that scattering rates for the in-plane transport 
\cite{7}  are dominated by the behavior measured in the nodal region, and are 
consistent with models in which the $c$-axis transport is dominated by physics 
from $(\pi,0)$ region \cite{8,9,10}. In the superconducting state, for points 
close to the nodal line, the widths continue their linear temperature dependence 
smoothly from the normal state. Further away from this nodal line, the 
scattering rates become temperature-independent in the superconducting state. 

The experimental studies reported in this paper were carried out on a Scienta 
electron spectrometer \cite{11}. In this instrument, the total spectral response 
may be measured as a function of angle and energy simultaneously. The instrument 
has an angular resolution of $0.1^\circ$ or better and, in the present studies, 
an energy resolution of the order of 10 meV. Photons were provided either by a 
resonance lamp or by a Normal Incidence Monochromator based at the National Synchrotron Light Source. In the photon energy range used, 15-21.22 eV, the momentum resolution is of the order of 0.005 \AA$^{-1}$. Samples of optimally doped ($T_C=91$ K) $Bi_2Sr_2CaCu_2O_{8+\delta}$, produced by the floating zone method \cite{12}, were mounted on a liquid He cryostat and cleaved {\it in-situ} in the UHV chamber with base pressure $3\times 10^{-9}$ Pa. During the recording of each spectrum, the temperature was measured using a silicon sensor mounted close to the sample. The self-energy may be determined either from energy distribution curves (EDC) or from momentum distribution curves (MDC). The EDC represents a measure of the intensity as a function of binding energy at constant momentum and the MDC represents a measure of the intensity as a function of momentum at constant binding energy.

In Fig.\ref{fig:1}(a) we show the Fermi surface determined in the 
superconducting state. This Fermi surface is obtained directly by measuring the 
peak positions in MDCs. We note that the area enclosed by this Fermi surface is 
consistent with the hole concentration in this optimally doped material. In the 
rest of the figure, we show representative spectra for line (2) in the normal 
state (b) and the superconducting state (c) and the same for line (4), closer to 
the $(\pi,0)$ region in panels (d) and (e). These spectra represent the measured 
photoemission intensities as a function of binding energy and parallel momentum. 
On line (2) the normal state shows a reasonably well defined dispersing band in 
the normal state, sharpening as one moves into the superconducting state. On 
line (4) the normal state is less well defined but the superconducting state is 
characterized by a sharp peak appearing close to the Fermi level. It should be 
noted that this peak actually shows considerable dispersion, as evident in 
panel (e). 

In Fig.\ref{fig:2} we show in the left panel EDCs for temperatures above and 
below the transition temperature, for three points on the Fermi surface 
indicated in Fig.\ref{fig:1}(a). In the normal state, peaks become progressively 
more ill-defined as we move closer to the $(\pi,0)$ region. However, the transition to the superconducting state is marked by the appearance 
of a sharp peak in the spectra. The anisotropy of the superconducting gap can 
clearly be seen with the largest gap in the vicinity of the $(\pi,0)$ point, as 
found in earlier studies \cite{2}. In the right panel of the figure, we show 
MDCs measured at the corresponding lines. The measurements in the normal state 
are recorded at the Fermi energy $(\omega=0)$. Below $T_C$, we 
show two measurements, the intensity recorded at $(\omega=0)$, and the 
intensity recorded at the leading edge, $\omega=-|\Delta({\bf k}_{F})|$. It is 
obvious that in MDCs, well-defined peaks exist in the normal state even in the 
vicinity of the $(\pi,0)$ point. For this reason, we focus on MDCs in our 
analysis of temperature dependence of the spectral width.

In Fig.\ref{fig:3}, we show the measured momentum widths as a function of 
temperature, for different points around the Fermi surface \cite{lines}. Measurements are 
made at the Fermi level in the normal state and at the leading edge in the 
superconducting state. The temperature dependence in the normal state is linear 
for most of the Fermi surface from the $(\pi,\pi)$ direction out towards the 
$(\pi,0)$ point. Furthermore, the temperature slope is similar over most of the 
same region. However, there is a marked increase in the temperature-independent 
offset as one moves away from the node.  Indeed we may describe the temperature 
dependence of momentum widths as $\Delta k({\bf k}_F,T)=a({\bf k}_F)+bT$, where 
$a({\bf k}_F)$ is momentum dependent but temperature independent and $b$, the 
slope, is approximately momentum independent. In the immediate vicinity of the 
$(\pi,0)$ point, it is possible that the slope has leveled off, leaving 
no temperature dependence in the normal state. The measurements from this latter 
region may be influenced by the strong $k$-dependence of matrix-elements near 
the $(\pi,0)$ point, and additionally complicated by the presence of the 
umklapps (see bottom right panel of Fig.\ref{fig:2}, for example). Therefore, 
uncertainties are largest in this region. In the superconducting state, the 
momentum widths appear to saturate as one moves away from the nodal line.  This 
is consistent with the emergence of a sharp peak in the EDCs, which has a 
temperature-independent width \cite{13}. 

The product of the momentum widths and the Fermi velocities provide the 
scattering rates or inverse lifetimes. In Fig.\ref{fig:4}(a), we show the 
measured velocities as a function of position around the Fermi surface. 
Velocities were obtained from dispersions deduced from MDCs. The low energy part 
in such a dispersion ($-50$ meV $<\omega<0$, for the normal state, and $-50$ 
meV$<\omega<-|\Delta({\bf k}_F)|$, for the superconductiong state) is then 
fitted by a straight line, with the slope representing the velocity. The 
velocity is shown both for the normal state, $v_N$, where it appears to be 
nearly constant over a large fraction of the Fermi surface, and for the 
superconducting state, $v_{SC}$ \cite{14}. The ratio of the velocities 
$v_N/v_{SC}$ is also shown. Note that due to the change in velocity, the 
scattering rates are significantly reduced below $T_C$ for points away from the 
node. 

Finally, in Fig.\ref{fig:4}(b) we show the single-particle scattering rates at 
different points on the Fermi surface for two temperatures, 100 K and 300 K. 
These scattering rates are obtained by multiplying the momentum widths in figure 
\ref{fig:3} by the normal-state velocities indicated in figure \ref{fig:4}(a). 

Now we focus on different aspects of the results presented here for the normal 
state. From the temperature and ${\bf k}_F$ dependence of the single-particle 
scattering rate (for $\omega=0)$, it is possible to calculate the conductivity 
in the normal state. In a simple Drude-type model, the conductivity is 
proportional to the integral of $k_Fl$ over the Fermi surface, where $k_F$ is 
the Fermi wave vector and $l=1/\Delta k$ is the mean free path. However, the 
observation in Fig.\ref{fig:3} that $\Delta k$ has a negligible zero-temperature 
offset only along the node, $a({\bf k}_F)\approx0$, shows that a simple 
integration would give an incorrect result for resistivity, the latter acquiring 
a significant $T$-independent term. This means that either the nodal excitations 
play a special role in the normal state transport, or single-particle scattering 
rates differ significantly from transport rates. That the normal state transport 
might be dominated by the behavior found in the nodal region, is not 
unreasonable if one considers the underlying antiferromagnetic structure of 
these materials. Along the diagonal direction, the spins on neighboring copper 
sites are ferromagnetically aligned. Along the copper oxygen bond direction 
transport will be frustrated by the antiferromagnetic alignment of the spins on 
neighboring copper sites. However, it is also true that transport discriminates 
scattering events, emphasizing large momentum transfers (small-angle events do 
not degrade measured currents). For example, recent thermal transport 
measurements on YBCO have indicated a sharp increase in the mean free path below 
$T_C$ \cite{15}, a behavior different from that found in ARPES for nodal 
excitations \cite{5}. When the system enters the superconducting state, the 
phase space for large momentum (angle) transfers collapses and the nodal 
excitations decay only through the small-angle events. Evidently, the scattering 
rates measured in thermal transport will be affected by the transition much more 
than the single-particle scattering rates measured in ARPES.

Recently, Abrahams and Varma have suggested that the temperature-and 
energy-independent term, $a({\bf k}_F)$, represents the scattering on static 
impurities, placed between the $CuO_2$ planes \cite{16}. Such impurities would 
indeed give rise to small-angle scattering, contributing only to single-particle 
scattering rates and producing a negligible effect on the resistivity. The 
strong momentum dependence of $a({\bf k}_F)$ is then explained by a variation in 
the density of states which is available for small-angle scattering at the 
corresponding momentum. In this picture, the only term relevant for the normal 
state transport is the temperature- and energy-dependent, but momentum 
independent marginal Fermi liquid self-energy. 

In a different approach to explain the linearity in the measured resistivity, 
Ioffe and Millis \cite{9} have suggested that the single-particle scattering 
rates should contain a temperature-independent term with $sin^2(2\phi)$ 
dependence, where $\phi$ is defined as in Fig.\ref{fig:4}. However, the present 
data show approximately linear dependence for small $\phi$ and more importantly, 
the linear temperature and energy dependence for the nodal direction \cite{5}, 
as opposed to the quadratic dependence of a Fermi liquid, used in their model.

Consideration of the anisotropy of the $c$-axis hopping integral has lead 
several authors to propose that the anomalous $c$-axis transport in these 
materials is dominated by scattering rates in the vicinity of the $(\pi,0)$ 
point \cite{8,9,10}. The $c$-axis resistivity for the optimally doped material 
is approximately constant over the range from 250 K down to 150 K at which point 
there is an increase before a rapid drop to zero at $T_C$ \cite{17}. The experimental points in 
Fig.\ref{fig:3} corresponding to the $(\pi,0)$ region are consistent with  this 
temperature dependence. Any integration over the Fermi surface, even if weighted 
by matrix elements, would give rise to a linear term in the $c$-axis 
resistivity. Our preliminary results on highly overdoped samples $(T_C\approx50 
$ K), show that the correspondence between the behavior in $(\pi,0)$ region and 
$c$-axis resistivity may indeed be generic and exist trough a wide range of 
doping levels \cite{18}. In the latter case, we have detected greater 
temperature dependence in $(\pi,0)$ region than in optimally doped samples, in 
agreement with the $c$-axis resistivity becoming more metallic in the overdoped 
regime. This correspondence is a strong indication that the anomalous $c$-axis 
transport may be a consequence of the in-plane physics in the $(\pi,0)$ region. 

In summary, the present study has shown that in the normal state, the linear 
temperature dependence observed for the imaginary part of the self-energy 
extends over at least 70\% of the Fermi surface. The scattering rates are highly 
anisotropic with a minimum along the nodal direction. 

The authors would like to acknowledge useful discussions with V. J. Emery, B. O. 
Wells, C. M. Varma, E. Abrahams, and G. Sawatzky.  The work was supported in 
part by the Department of Energy under contract number DE-AC02-98CH10886 and in 
part by the New Energy and Industrial Technology Development Organization 
(NEDO).

\begin{figure}
\caption{
(a) Fermi surface of the optimally doped Bi 2212, measured in the 
superconductiong state. Indicated are the lines (1) to (5) on which the 
temperature dependence is measured. Typical spectra are shown for line (2) in 
the normal (b) and superconducting state (c), as well as for line (4), in the 
normal (d) and superconducting state (e).
}
\label{fig:1}
\end{figure}

\begin{figure}
\centerline{\epsfxsize=7cm\epsfbox{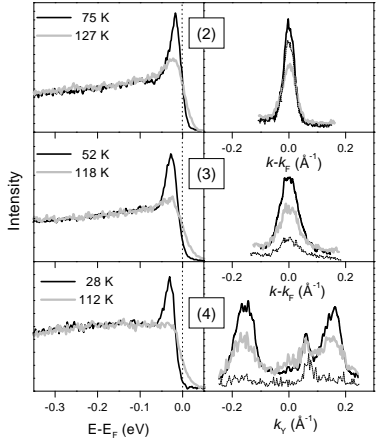}}
\caption{
EDCs (left panel) and MDCs (right panel) for lines (2), (3) and (4) from Fig. 
1(a), in the normal (gray lines) and superconductiong state (black lines). In 
the superconducting state, the MDCs are measured at the Fermi level (dashed 
lines) and at the leading edge (solid lines).
}
\label{fig:2}
\end{figure}

\begin{figure}
\centerline{\epsfxsize=7cm\epsfbox{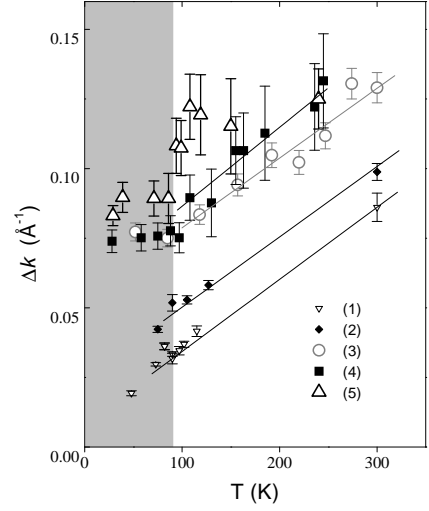}}
\caption{
Momentum widths as a function of temperature for different positions on the 
Fermi surface, obtained by fitting the MDCs with Lorentzian lineshapes. Widths 
are measured at the Fermi level and at the leading edge, in the normal and in 
superconductiong (gray region) state, respectively.
}
\label{fig:3}
\end{figure}

\begin{figure}
\centerline{\epsfxsize=6cm\epsfbox{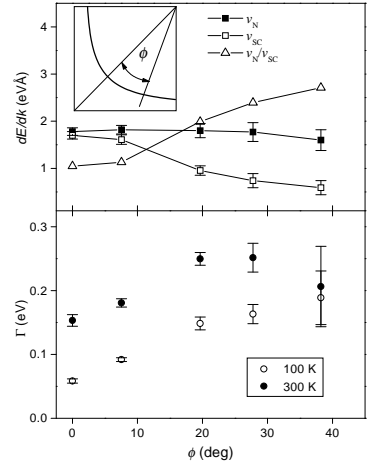}}
\caption{
(a) Velocities along the lines from Fig. 1(a) in the normal (solid squares) and 
superconducting (open squares) state as a function of the angle $\phi$, defined 
in the inset (see text for details). Ratio between the normal state and 
superconductiong state velocities is also shown (open triangles).
(b) Normal state scattering rates as a function of $\phi$, obtained by 
multiplying momentum widths from Fig. 3 with normal state velocities.
}
\label{fig:4}
\end{figure}

\begin{references}
\bibitem{lines} The lines (2) to (5) are not perpendicular to the Fermi surface. However, since the discrepancy is small, and since $\Delta k<<k_{F}$, it is straightforward to extract perpendicular components of momentum widths and velocities (shown in Fig. 3 and Fig. 4(a)).
\end{references}
\end{document}